\documentclass{pasj00}
\draft

\begin{document}
\SetRunningHead{T. Morioka et al.}{Star-forming galaxies at $z=0.24$}
\Received{yyyy/mm/dd}
\Accepted{yyyy/mm/dd}

\title{STAR-FORMING GALAXIES AT $z \approx 0.24$ IN THE SUBARU DEEP FIELD 
AND THE SLOAN DIGITAL SKY SURVEY\altaffilmark{1}}

\author{Taichi \textsc{Morioka},\altaffilmark{2, 3}
        Aki \textsc{Nakajima},\altaffilmark{4}
        Yoshiaki \textsc{Taniguchi},\altaffilmark{5}
        Yasuhiro \textsc{Shioya},\altaffilmark{5}
        Takashi \textsc{Murayama},\altaffilmark{2} \& 
        Shunji S. \textsc{Sasaki}\altaffilmark{2,4}
        }

\email{shioya@cosmos.ehime-u.ac.jp}
\email{tani@cosmos.ehime-u.ac.jp}

\altaffiltext{1}{Based on data collected at 
	Subaru Telescope, which is operated by 
	the National Astronomical Observatory of Japan.}
\altaffiltext{2}{Astronomical Institute, Graduate School of Science,
        Tohoku University, Aramaki, Aoba, Sendai 980-8578}
\altaffiltext{3}{Nikkei Home Publishing, INC., Akasaka, Minato-ku, Tokyo 107-8406}
\altaffiltext{4}{Graduate School of Science and Engineering, Ehime University, 
        Bunkyo-cho, Matsuyama 790-8577}
\altaffiltext{5}{Research Center for Space and Cosmic Evolution, Ehime University, 
        Bunkyo-cho, Matsuyama 790-8577}


\maketitle

\begin{abstract}
We make a search for H$\alpha$ emitting galaxies at $z \approx 0.24$ in 
the Subaru Deep Field (SDF) using the archival data set obtained with the Subaru Telescope. 
We carefully select H$\alpha$ emitters in the narrowband filter $NB816$, using $B$, $V$, 
$R_{\rm c}$, $i^\prime$, and $z^\prime$ broad-band colors. 
We obtain a sample of 258 emitting galaxies with observed equivalent widths of 
(H$\alpha$+[N{\sc ii}]6548,6584) greater than 12 \AA.
We also analyze a sample of H$\alpha$ emitters taken from the Sloan Digital Sky Survey (SDSS) 
to constrain the luminous end of H$\alpha$ luminosity function. 
Using the same selection criteria as for the SDF, and after excluding AGNs, 
we obtain 317 H$\alpha$ emitting star-forming galaxies.
Combining these two samples of H$\alpha$ emitters found in both SDF and SDSS, 
we derive a H$\alpha$ luminosity function with best-fit Schechter function parameters 
of $\alpha = -1.31^{+0.17}_{-0.17}$, $\log \phi^{*} = -2.46^{+0.34}_{-0.40} {\rm Mpc^{-3}}$, $\log L^{*} = 
41.99^{+0.08}_{-0.07} {\rm ergs \; s^{-1}}$. 
An extinction-corrected H$\alpha$ luminosity density is 
$4.45^{+2.96}_{-1.75} \times 10^{39}\;{\rm ergs \; s^{-1} \; Mpc^{-3}}$.
Using the Kennicutt relation between the H$\alpha$ luminosity and star 
formation rate, the star formation rate density in the survey volume is 
estimated as $0.035^{+0.024}_{-0.014} ~ M_{\odot} \; {\rm yr^{-1} \; Mpc^{-3}}$.
The angular two-point correlation function of H$\alpha$ emitters over 875
$\rm arcmin^2$ at $z = 0.24$ is well fitted by 
a power-law form with $w(\theta) = 0.047^{+0.017}_{-0.013} \theta^{-0.66 \pm 0.08}$, 
corresponding to the correlation function of $\xi(r) = (r/{\rm 2.6^{+1.0}_{-0.8}Mpc})^{-1.66\pm0.08}$. 
The small correlation length of H$\alpha$ emitters may imply the weak clustering 
of active star-forming galaxies. 
\end{abstract}

\section{INTRODUCTION}
 
Since the cosmic star formation history in galaxies from high-redshift
to the present day is one of important measurements for the understanding of
both formation and evolution of galaxies, a number of observational
studies have been made for this decade (e.g., Madau et al. 1996;
Madau, Pozzetti, \& Dickinson 1998; Pettini et al 1998; Steidel et al. 1999;
Barger et al. 2000; Giavalisco et al. 2004; Dickinson et al. 2004;
Taniguchi et al. 2005; Bouwens \& Illingworth 2006; 
see also Hopkins 2004; Hopkins \& Beacom 2006). 
Although much interest has been addressed to
the star formation in higher-redshift universe, large programs of
spectroscopic surveys have provided important observations of
galaxies located in lower-redshift universe; e.g.,  Canada-France
Redshift Survey (CFRS: Lilly et al. 1995, 1996; Tresse \& Maddox 1998; Tresse et al. 2002),
the Sloan Digital Sky Survey (SDSS: York et al. 2000; Salim et al. 2007),
the VIMOS VLT Deep Survey (VVDS: Le Fevre et al. 2005; Tresse et al. 2007).
All these observations have provided us a rough trend of the
cosmic star formation history from high-$z$ to the present day: 
the star-formation rate density (SFRD, $\rho_{\rm SFR}$) rises from $z \simeq 0$ to $z \simeq 1$, 
is nearly constant between $z \sim 2$ and $z \sim 5$ and decline at $z > 5$. 
Recent observations by the {\it Galaxy Evolution Explorer} (GALEX) and 
the {\it Spitzer Space Telescope} have confirmed that $\rho_{\rm SFR}$ increases from 
$z \sim 0$ to $z \sim 1$ (e.g., Schiminovich et al. 2005; Le Floc'h et al. 2005). 

Remaining problems include the faint-end and luminous-end of the luminosity function 
and clustering properties of star-forming galaxies. 
For this purpose, deep imaging
observations  with a narrowband filter are highly useful because such surveys could probe
fainter star-forming galaxies by detecting their H$\alpha$ emission.
The H$\alpha$ luminosity is directly connected to the ionizing photon production rate. 
In particular, wide-field imagers on 8m class telescopes,
such as Suprime-Cam on the Subaru Telescope (Miyazaki et al. 2002)
are most suitable because of their superior observational capability. 

In this paper, we present our new search for faint H$\alpha$ emitting galaxies
at $z = 0.24$ in the Subaru Deep Field (SDF: Kashikawa et al. 2004). 
One of the uniqueness of the SDF is the field is its very wide field of view ($34^\prime \times 27^\prime$). 
Another uniqueness is that deep photometry in narrowband filters has been carried out. 
Since the SDF is one of the deepest fields observed so far by ground-based telescopes,
we can probe much fainter star-forming galaxies. 
Although the sky coverage of SDF is fairly wide,
it is difficult to sample a statistically significant number of bright H$\alpha$ emitters.
This makes it difficult to obtain a reliable H$\alpha$ luminosity function (H$\alpha$LF)
because of large ambiguity in the bright end of luminosity function.
Therefore, in this paper, we also use the SDSS data 
to probe the bright-end of the luminosity function of H$\alpha$ emitters 
in the same redshift range. 

We note that Ly et al. (2007, hereafter L07) discussed low and intermediate redshift NB704, NB711, 
NB816, and NB921 emitters in the SDF using the official photometric catalogs. 
Since their selection criteria are different from those of this work, 
it is hard to compare with our previous work (Shioya et al. 2008, hereafter S08). 
We therefore reanalyze the public catalog of the SDF. 
Details of the differences between our sample and the sample of L07 
are discussed in the following section. 

Throughout this paper, magnitudes are given in the AB system.
We adopt a flat universe with $\Omega_{\rm matter} = 0.3$,
$\Omega_{\Lambda} = 0.7$,
and $H_0 = 70 \; {\rm km \; s^{-1} \; Mpc^{-1}}$.

\section{H$\alpha$ EMITTERS AT $z \approx 0.24$ IN THE SDF}

\subsection{SDF Photometric Catalog}

The SDF project is a very deep optical imaging survey using the Suprime-Cam
(Miyazaki et al.2002) on the 8.2 m Subaru Telescope (Kaifu et al. 2000;
Iye et al. 2004) at Mauna Kea Observatories.
The SDF is located near the North Galactic Pole, being centered at $\alpha$(J2000) 
= $13^{\rm h} 24^{\rm m} 38^{\rm s}.9$ and $\delta$(J2000) = $+27^\circ
29^\prime 25^{\prime\prime}.9$.
Details of the SDF project are given in Kashikawa et al.(2004).
The SDF official photometric catalog
is obtained from the SDF site (http://step.mtk.nao.ac.jp/sdf/project/).
This official catalog contains 5 broadband ($B,V,R_{c},i^{\prime}$, and $z^{\prime}$)
and 2 narrowband ($NB816$ and $NB921$) photometric data.
In this work, we use an $i^\prime$-selected catalog with $2^{\prime \prime}$ diameter aperture photometry.
The PSF size in this catalog is 0.98 arcsec (Kashikawa et al. 2004).
The $3 \sigma$ limiting magnitudes, which are measured as $3 \sigma$ of the sky noise 
on a $2^{\prime \prime}$ diameter aperture, are $B=28.45$, $V=27.74$, $R_{c}=27.80$, 
$i^\prime=27.43$, $z'=26.62$, and $NB816=26.63$, respectively. 
Since the Galactic extinction is not corrected in the magnitudes in 
the official catalog, we applied the Galactic extinction correction of 
$E(B-V)=0.017$ (Schlegel et al. 1998). 
A photometric correction for each band is as follows: 
$A_B=0.067$, $A_V=0.052$, $A_{R_{\rm C}}=0.043$, $A_{i^\prime}=0.033$, 
$A_{z^\prime}=0.025$, and $A_{NB816}=0.030$. 

In this work, we use the narrowband filter, $NB816$, centered at 8150 \AA~
with the passband of $\Delta\lambda$(FWHM) = 120 \AA.
The central wavelength of this filter corresponds to a redshift of 0.24
for H$\alpha$ emission. Therefore, together with the broadband filter data,
we can carry out a deep search for H$\alpha$ emitters at $z \approx$ 0.24 in SDF.

\subsection{Selection of $NB816$-Excess Objects}

In order to select $NB816$-excess objects, 
it would be desirable to have a frequency-matched continuum. 
Since the effective frequency of the $NB816$ filter (367.8 THz) is different either from 
those of $i^\prime$ (349.9 THz) and $z^\prime$ (333.6 THz) filters, we newly make a 
frequency-matched continuum, ``$iz$ continuum'', for each object using the following
linear combination; $iz = 0.57 f_{i^\prime}+0.43 f_{z^\prime}$ 
where $f_{i^\prime}$ and $f_{z^\prime}$ are the $i^\prime$- and $z^\prime$-band
flux densities, respectively.
Its 3 $\sigma$ limiting magnitude is $iz \simeq 27.0$ in a $2^{\prime \prime}$ 
diameter aperture.
This enables us to more precisely estimate the continuum magnitude at the same 
effective wavelength as that of the $NB816$ filter. 
However, in SDF public catalog, the median of $iz-NB816$ is 0.08 for objects with $21 < NB816 < 24$. 
We therefore corrected the $NB816$ magnitude; $NB816 = NB816$(pub)+0.08, 
where $NB816$(pub) is magnitude of $NB816$ in SDF public catalog.
From now on, we use this $NB816$ magnitude. 
We note that the median of $i^\prime z^\prime - NB816$ measured by L07 (0.1) 
is slightly larger than our measurement. 
We consider that the different definition of $iz$ continuum makes the difference of 
the median of $iz-NB816$. 

\begin{figure}
\begin{center}
\FigureFile(150mm,150mm){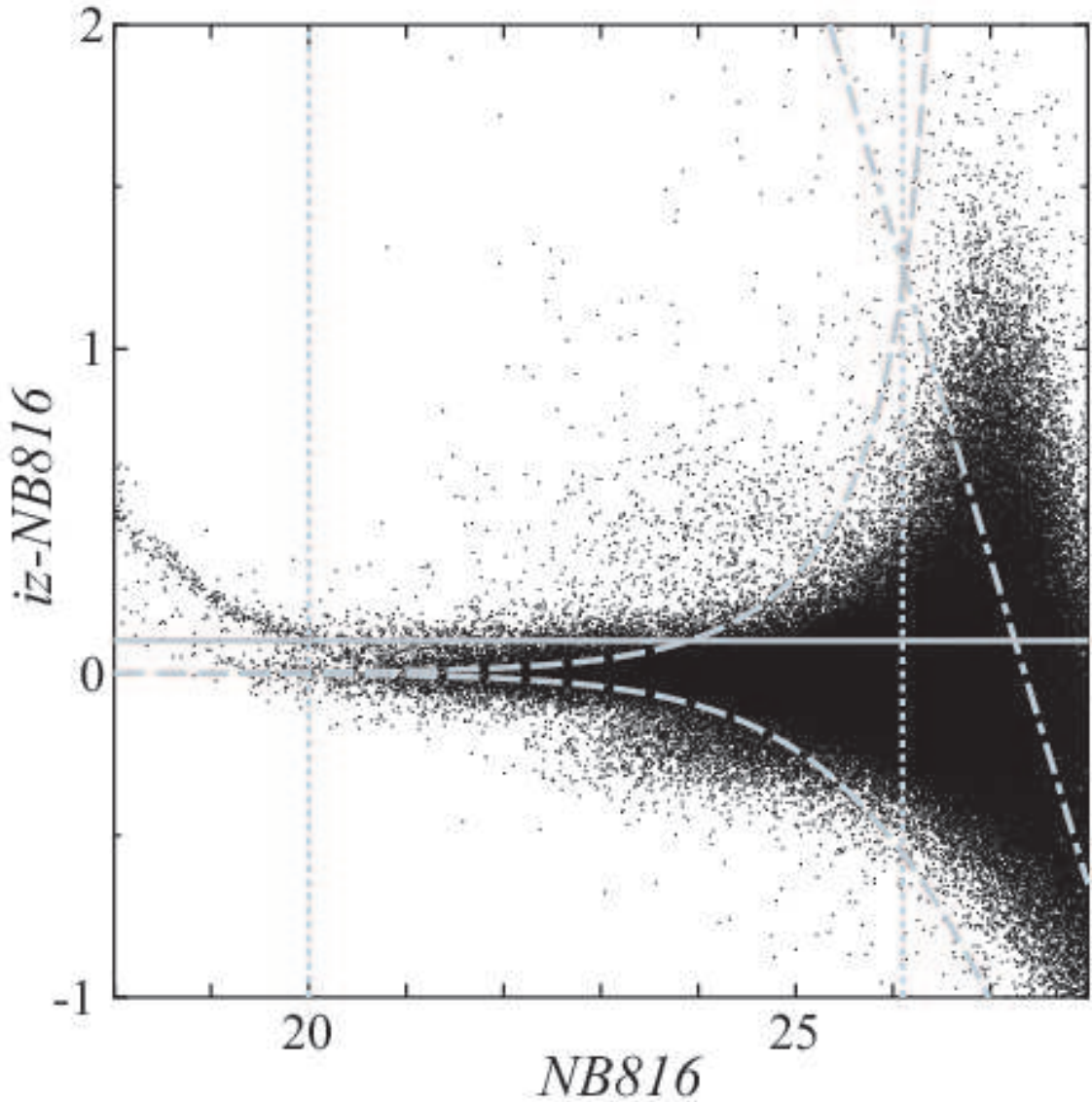}
\end{center}
\caption{Diagram between $iz-NB816$ and $NB816$. 
The gray horizontal solid line shows our selection criterion of 
the NB816-excess objects, that corresponds to $iz-NB816 = 0.1$.
The gray vertical dotted lines corresponds to $NB816 = 20$ and $26.1$.
The gray dashed lines show the distribution of $3\sigma$ error. 
The gray dot-dashed line shows the $3 \sigma$ limiting magnitude of $iz$. 
We note that this diagram is not comparable with the Figure 1 of L07 
(see section 2.2). 
\label{NBexcess}}
\end{figure}

Following the manner of previous surveys using a narrowband filter 
(Fujita et al. 2003, hereafter F03; L07; S08) 
and taking the scatter in the $iz - NB816$ 
color and our survey depth into account, candidate line-emitting objects are 
selected with the criteria of $iz - NB816 \geq \max(0.1,3\sigma_{iz-NB816})$ and 
$20 \le NB816 \le 26.1$. 
In order to avoid the influence of saturation of brighter objects,
we have adopted a criterion of $NB816 > 20$. 
We note that $NB816 = 26.1$ is determined from intersection point of
$NB816 + 3\sigma_{iz-NB816}= iz_{3\sigma}$. 
We compute the $3\sigma$ of the $iz - NB816$  color as
$3\sigma_{iz - NB816} = -2.5\log
(1-\sqrt{(f_{3\sigma_{NB816}})^2+(f_{3\sigma_{iz}})^2}/f_{NB816})$.
We also note that $iz - NB816 = 0.1$ corresponds to $EW_{\rm{obs}} \approx 12$ \AA.
These criteria are shown by the vertical dotted lines, the horizontal solid line, 
and the dashed curve in Figure \ref{NBexcess}.
We then find 2072 sources that satisfy the above criteria. 
The number of our emitter candidates is larger than that of L07. 
We discuss the differences between our sample and the sample of L07 
in section 2.5. 

\subsection{Selection of H$\alpha$ Emitters at $z\approx0.24$}

A narrowband survey of emission-line galaxies can potentially detect galaxies with 
different emission lines at different redshifts. 
The emission lines that can be detected in the $NB816$ passband are 
H$\alpha$+[N{\sc ii}], H$\beta$, [O {\sc iii}] $\lambda \lambda$4959, 5007, 
[O {\sc ii}] $\lambda$3727, [S {\sc ii}] $\lambda \lambda$ 6717, 6731, and Ly$\alpha$. 
Since the flux ratio of [S {\sc ii}]$\lambda \lambda$ 6717, 6731/(H$\alpha$ + 
[N {\sc ii}] $\lambda\lambda$ 6548,6584) is low in general ($\sim 0.2$: 
Ho, Fillippenko, \& Sargent 1997; see also Murayama \& Taniguchi 1998; Nagao et al. 2006), 
we do not focus on these lines in this paper. 
As for the Ly$\alpha$ emitter search in SDF, see, e.g., Taniguchi et al. (2005), 
Shimasaku et al. (2006), and Kashikawa et al. (2006).
In Table \ref{Ha:tab:cover} we show the different redshift coverage for each line.

\begin{figure}
\begin{center}
\FigureFile(80mm,150mm){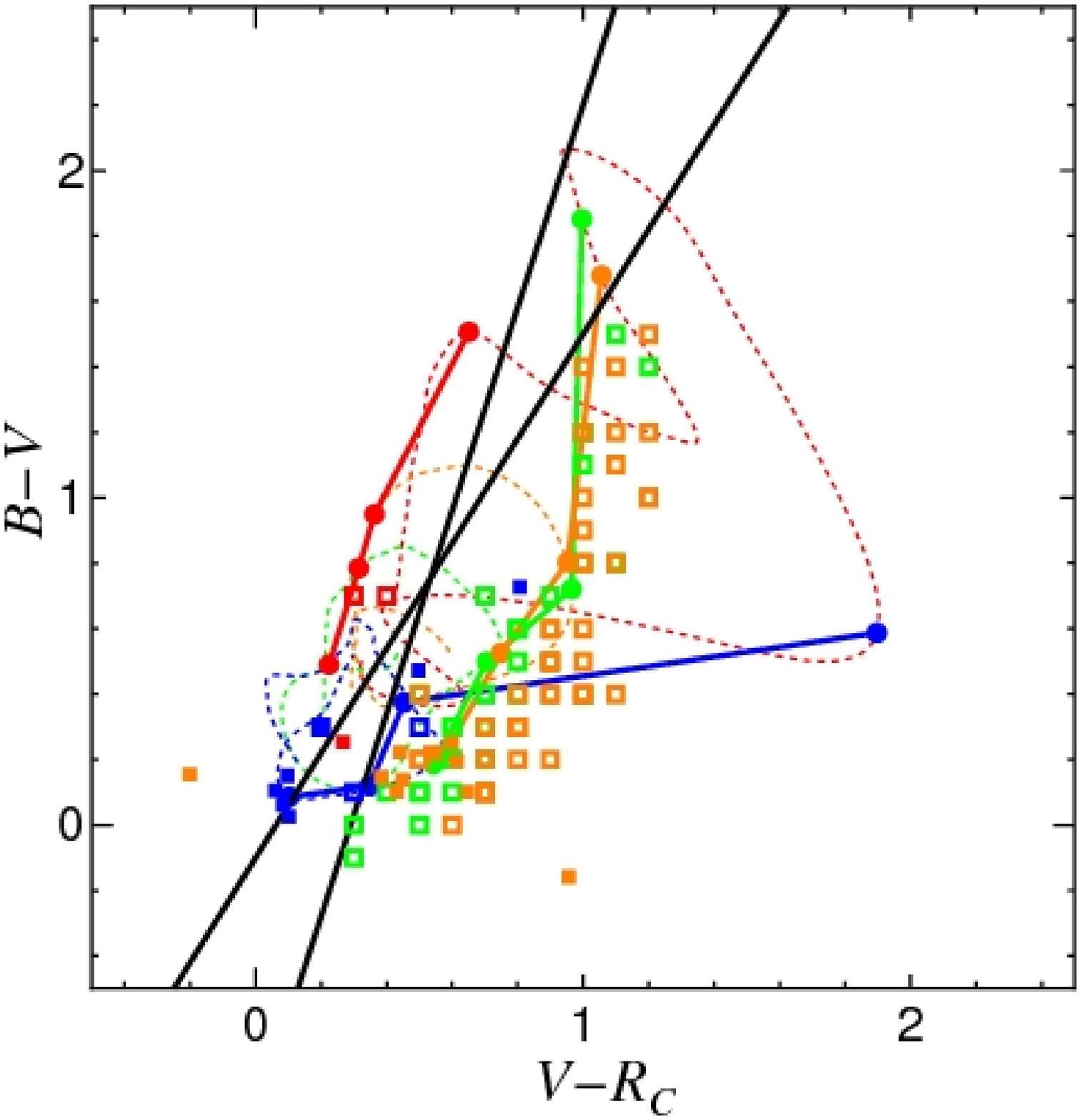}

\FigureFile(80mm,150mm){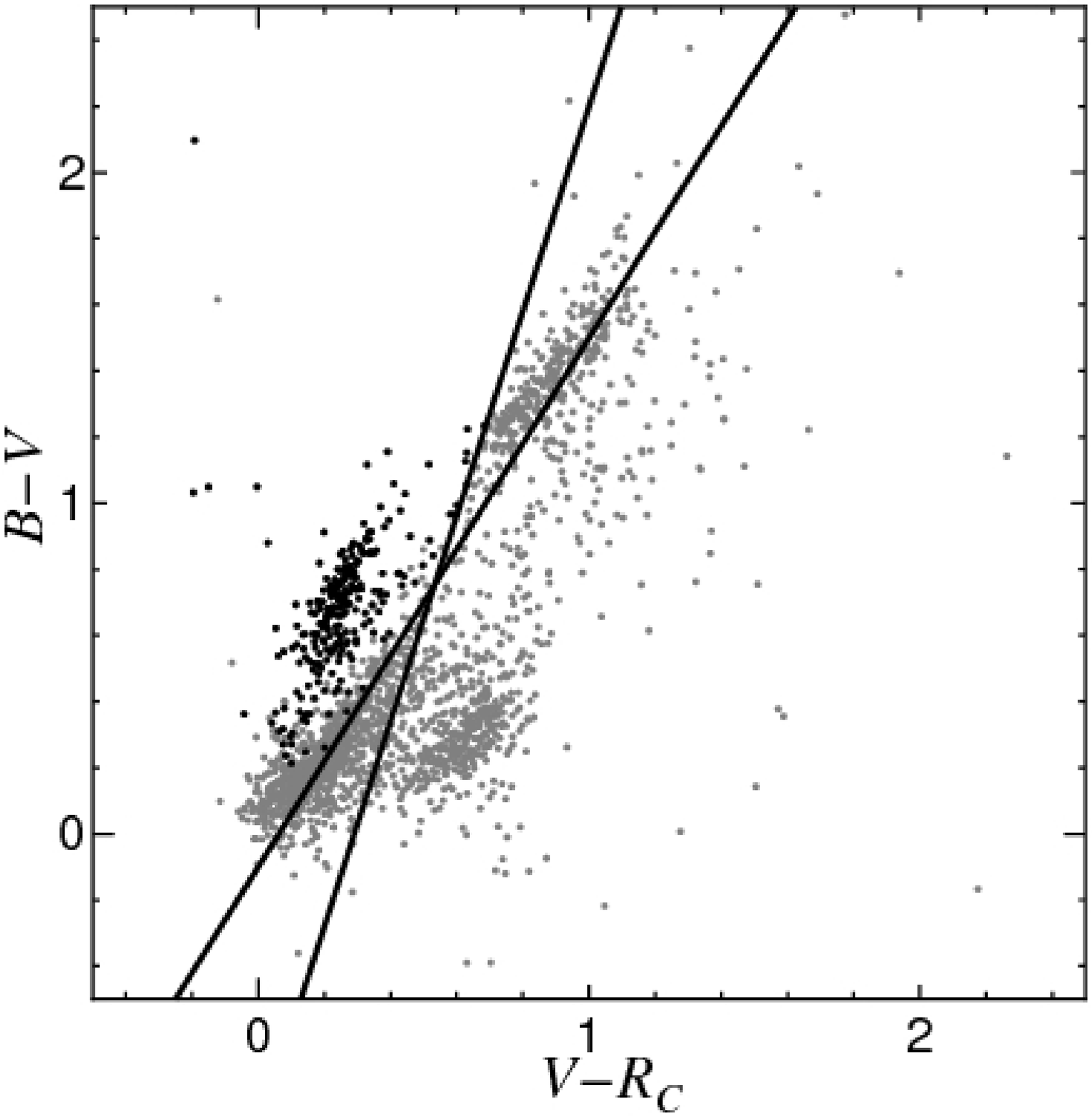}
\end{center}
\caption{
Diagrams of $B-V$ vs. $V-R_{\rm C}$. 
(Top) Colors of model galaxies (CWW) from $z=0$ to $z=3$ are shown with 
dotted lines: red, orange, green, and blue lines show the loci of E, Sbc, Scd, and Irr 
galaxies, respectively. 
Colors of $z=0.24$, $0.64$, $0.68$, and $1.18$
(for H$\alpha$, [O {\sc iii}], H$\beta$, and [O {\sc ii}] emitters, respectively) 
are shown with red, orange, green, and blue lines, respectively. 
Galaxies in the GOODS-N (Cowie et al. 2004) with redshifts corresponding to H$\alpha$ 
emitters, [O{\sc iii}] emitters, H$\beta$ emitters, and [O{\sc ii}] emitters are shown 
as red, orange, green, and blue open squares, respectively. 
Galaxies in the SDF (L07) with redshifts corresponding to H$\alpha$ 
emitters, [O{\sc iii}] emitters, and [O{\sc ii}] emitters are shown 
as red, orange, and blue filled squares, respectively. 
Solid lines show $(B-V) = 1.6(V-R_{\rm C})-0.1$ and $(B-V) = 3.1(V-R_{\rm C})-0.9$. 
H$\alpha$ emitters are located on the left side of the two lines. 
(Bottom) Plot of $B-V$ vs. $V-R_{c}$ for 2072 emission line candidates (gray filled circles) 
and 258 H$\alpha$ emitters (black filled circles).
\label{color1}}
\end{figure}

\begin{figure}
\begin{center}
\FigureFile(80mm,150mm){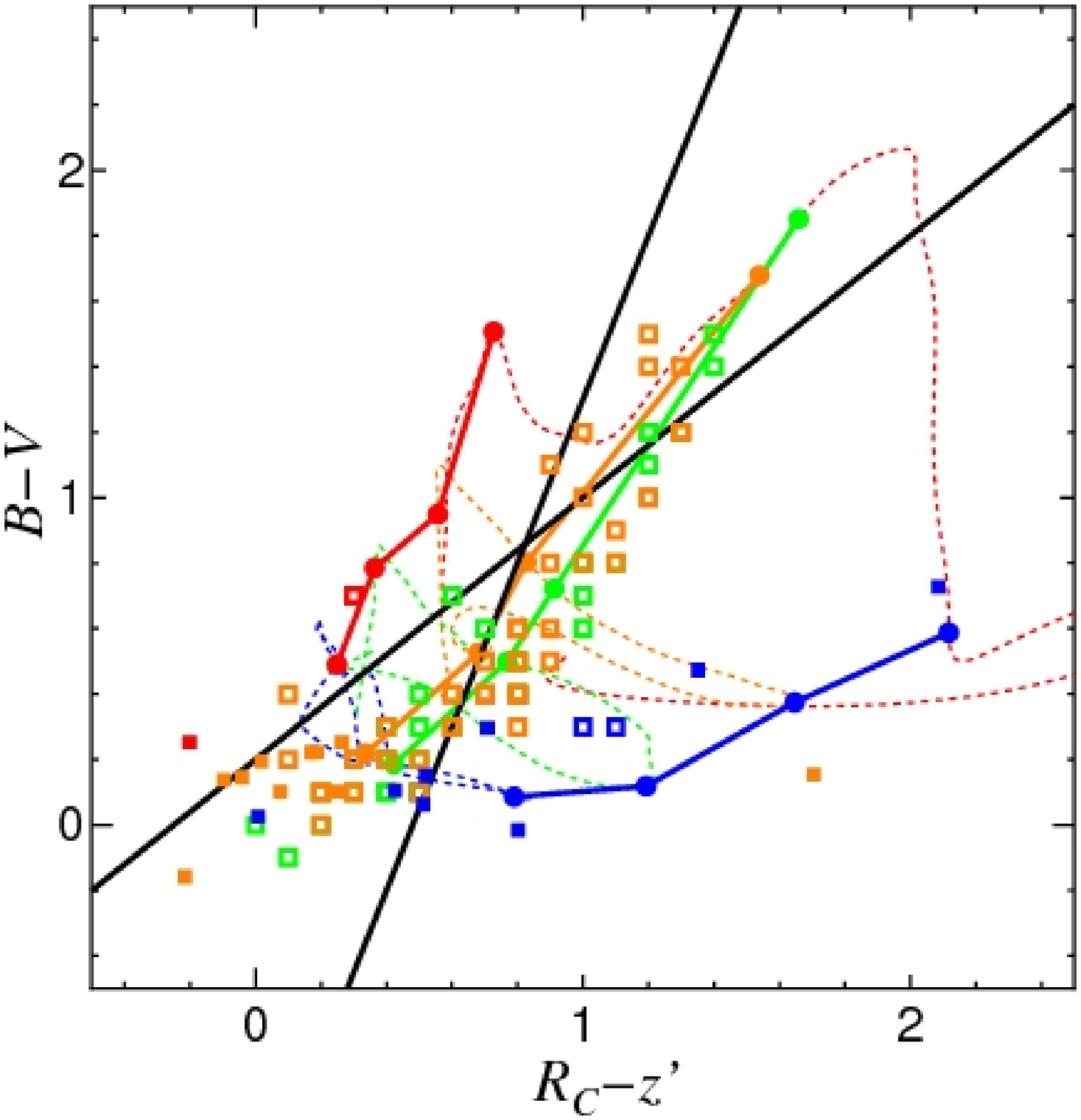}

\FigureFile(80mm,150mm){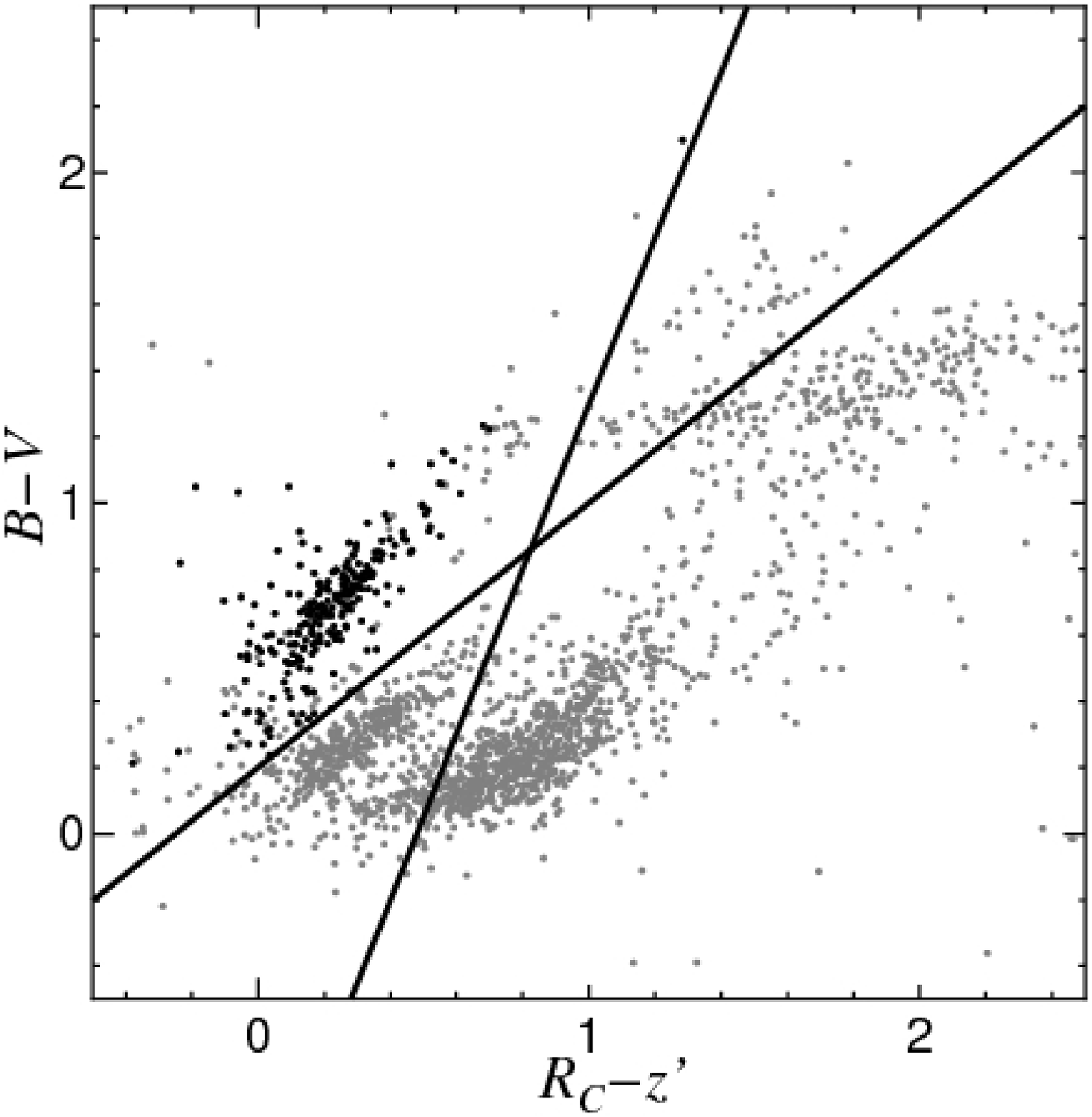}
\end{center}
\caption{
Diagrams of $B-V$ vs. $V-z^\prime$. 
(Top) Colors of model galaxies (CWW) from $z=0$ to $z=3$ are shown with 
dotted lines: red, orange, green, and blue lines show the loci of E, Sbc, Scd, and Irr 
galaxies, respectively. 
Colors of $z=0.24$, $0.64$, $0.68$, and $1.18$
(for H$\alpha$, [O {\sc iii}], H$\beta$, and [O {\sc ii}] emitters, respectively) 
are shown with red, orange, green, and blue lines, respectively. 
Galaxies in the GOODS-N (Cowie et al. 2004) with redshifts corresponding to H$\alpha$ 
emitters, [O{\sc iii}] emitters, H$\beta$ emitters, and [O{\sc ii}] emitters are shown 
as red, orange, green, and blue open squares, respectively. 
Galaxies in the SDF (L07) with redshifts corresponding to H$\alpha$ 
emitters, [O{\sc iii}] emitters, and [O{\sc ii}] emitters are shown 
as red, orange, and blue filled squares, respectively. 
Solid lines show $(B-V) = 0.8(R_{\rm C}-z^\prime)+0.2$ and $(B-V) = 2.5(R_{\rm C}-z^\prime)-1.2$. 
H$\alpha$ emitters are located on the left side of the two lines. 
(Bottom) Plot of $B-V$ vs. $V-R_{c}$ for 2072 emission line candidates (gray filled circles) 
and 258 H$\alpha$ emitters (black filled circles).
\label{color2}}
\end{figure}

In order to distinguish H$\alpha$ emitters at $z \approx 0.24$ from
emission-line objects at other redshifts,
we investigate their broad-band color properties. 
Specifically, we compare observed colors of our 2072 emitters with model 
ones that are estimated by using 
the model spectral energy distribution derived by Coleman, Wu, \& Weedman (1980).
In Figures \ref{color1} \& \ref{color2}, 
we show the $B-V$ vs. $V-R_{c}$ and 
$B-V$ vs. $R_{c} - z^\prime$ color-color diagram of the 2072 sources 
and the loci of model galaxies.
We find three clumps which are considered to correspond to the H$\alpha$+[N{\sc ii}], 
[O{\sc iii}]+H$\beta$, and [O{\sc ii}] emitters. 
We note that another clump at $B-V \sim 1.4$ and $V-R_{\rm C} \sim 0.8$ in Figure \ref{color1} 
is considered to be late-type stars. 
Then we find that H$\alpha$ emitters at $z\approx0.24$ can be selected by
adopting the following four criteria;
(1) $(B-V) > 1.6(V-R_{c}) - 0.1$,
(2) $(B-V) > 3.1(V-R_{c}) - 0.9$,
(3) $(B-V) > 0.8(R_{c}-z^\prime) + 0.2$, and 
(4) $(B-V) > 2.5(R_{c}-z^\prime) - 1.2$. 
Since all H$\alpha$+[N{\sc ii}] emitter candidates selected above are brighter than the limiting magnitude 
in each band, they are considered not to be Lyman $\alpha$ emitters. 
We can clearly distinguish H$\alpha$ emitters from [O {\sc iii}] or H$\beta$ emitters 
using the first and second criteria. 
We can also distinguish H$\alpha$+[N{\sc ii}] emitters from [O {\sc ii}] emitters 
using the third and fourth criteria. 
To investigate how our selection criteria suffer from 
contaminations of galaxies at different redshifts, we plot 
the colors of galaxies with a spectroscopic redshift 
[specifically, 2 H$\alpha$ emitters, 47 [O{\sc iii}] emitters, 36 H$\beta$ emitters
and 3 [O{\sc ii}] emitters presented in Cowie et al. (2004), and
1 Ha emitter, 10 [O{\sc iii}] emitters and 8 [O{\sc ii}] emitters presented
in L07] in Figures 2 and 3. There is no contamination
at least concerning this spectroscopic sample, which strongly
justify our selection criteria.These criteria give us a sample of 258 H$\alpha$+[N{\sc ii}] emitting galaxy candidates.
The number of our H$\alpha$ emitter candidates is larger than that of L07. 
We discuss the differences between our sample and the sample of L07 
in section 2.5. 

\subsection{H$\alpha$ Luminosity}

In order to obtain the H$\alpha$ luminosity for each source, 
we correct for the presence of [N {\sc ii}] lines.
Further, we also apply a mean internal extinction
correction to each object. For these two corrections,
we have adopted the flux ratio of
$f(\textrm{H}\alpha) / f([\textrm{N {\sc ii}}] \lambda\lambda 6548,6584) = 2.3$ 
(obtained by Kennicutt 1992; Gallego et al. 1997; used by Yan et al. 1999; 
Iwamuro et al. 2000; F03) and $A_{\rm{H}\alpha}=1$ (Kennicutt 1983; Niklas et al. 1997). 
We also apply a statistical correction (21\%; the average value of flux decrease
due to the filter transmission) to measured flux because the filter transmission
function is not square in shape (F03; L07; S08).
Note that this value is slightly different from the value (28\%) used in F03 and L07 
and re-evaluated by using the latest response function in S08. 
The H$\alpha$ flux is given by:
\begin{eqnarray}
f_{\textrm{cor}}(\textrm{H}\alpha) = f(\textrm{H}\alpha + [\textrm{N {\sc ii}}]) \times
\frac{f(\textrm{H}\alpha)}{f(\textrm{H}\alpha + [\textrm{N {\sc ii}}])} \times 
10^{0.4A_{{\rm H} \alpha}} \times 1.21
\end{eqnarray}
We calculate $f({\rm H+[N}${\sc ii}]) from the total magnitude;
note that the aperture size is determined from the $i^{\prime}$-band image.
Finally the H$\alpha$ luminosity is given by $L(\textrm{H}\alpha) = 
4\pi d_{\rm{L}}^2 f_{\rm{cor}}(\rm{H}\alpha)$
using the redshift of the line at the center of the filter $z=0.242$ and
the luminosity distance, $d_{\rm L}$. 
Although the derived H$\alpha$ luminosity ranges from $10^{39.4}$ to $10^{41.6}$ 
$\rm erg \; s^{-1}$ (Figure \ref{LF}), we consider our H$\alpha$ luminosity function 
is complete between $39.8 < \log L({\rm H}\alpha) < 40.8$. 
The luminosity, $\log L({\rm H}\alpha) = 40.8$, corresponds to 
that of the H$\alpha$ emitters with $iz-NB816 = 0.1$ and $NB816=20$. 

\subsection{Difference between our sample and that of Ly et al. (2007)}

We summarize here the difference of sample selection between ours and L07. 
L07 used $NB816$-selected catalog and 
selected H$\alpha$ emitters using the following criteria: 
(1) $iz-NB816 > 0.25$, 
(2) $B-V > 2(R_{\rm C}-i^\prime)-0.1$, and 
(3) $R_{\rm C}-i^\prime < 0.45$. 
We also note that L07 did not correct NB816 magnitude to set the median of 
$iz-NB816$ to zero. 

First, we compare our criterion for emission-line galaxies with those of L07. 
Since the median of $iz-NB816$ was 0.1 in L07, 
they selected galaxies with $EW_{\rm obs} > 18$\AA~ using the criterion (1). 
On the other hand, we select galaxies with $iz-NB816 > 0.1$. 
Taking account of the median of $iz-NB816=0$, 
we select emission-line galaxies with $EW_{\rm obs} > 12$ \AA. 
Basically, the emission-line galaxy candidates in L07 are 
included our emission-line galaxy candidates. 
The number of galaxies with $iz-NB816 > 0.15$ is 1388. 

\begin{figure}
\begin{center}
\FigureFile(80mm,150mm){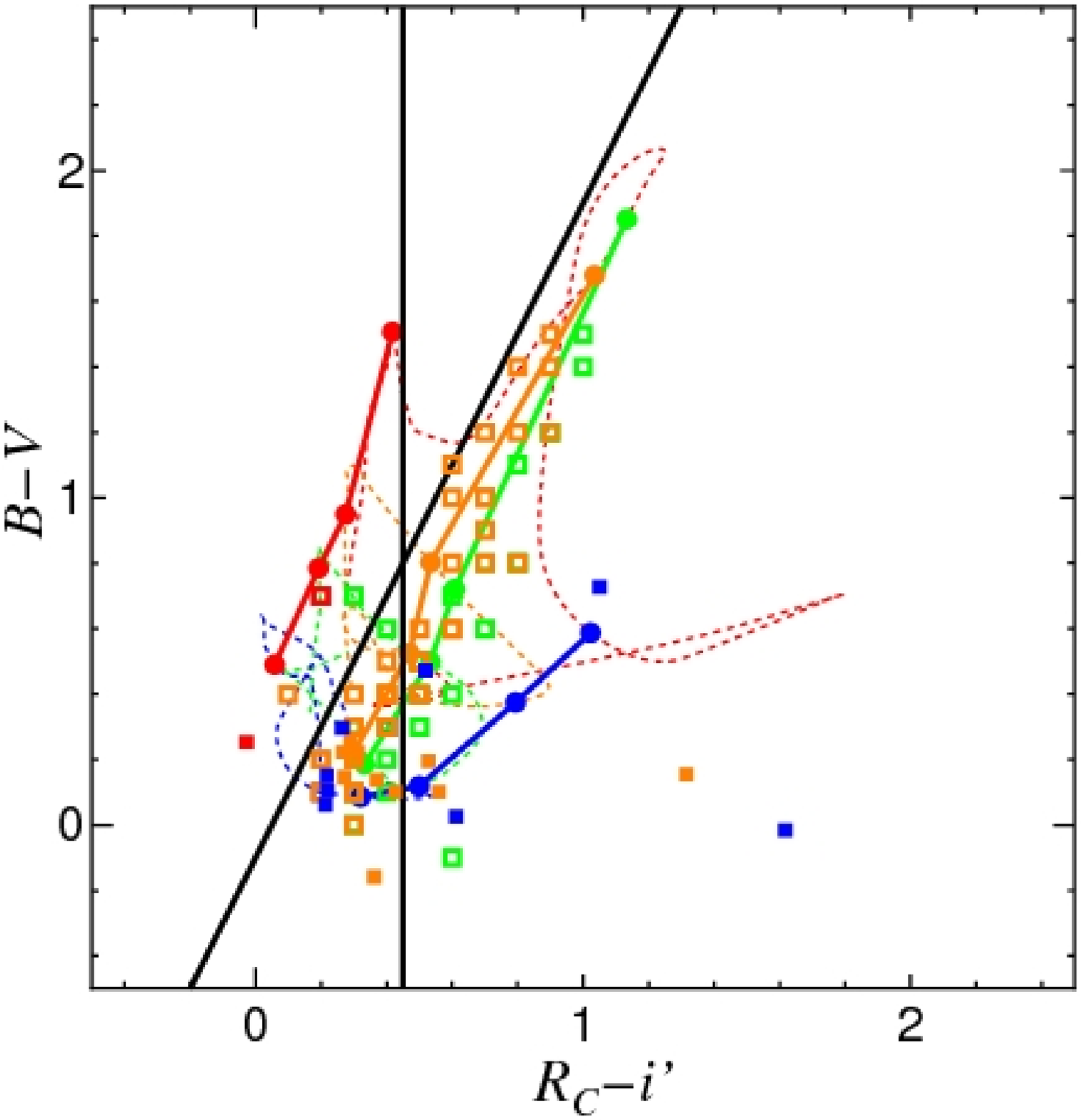}

\FigureFile(80mm,150mm){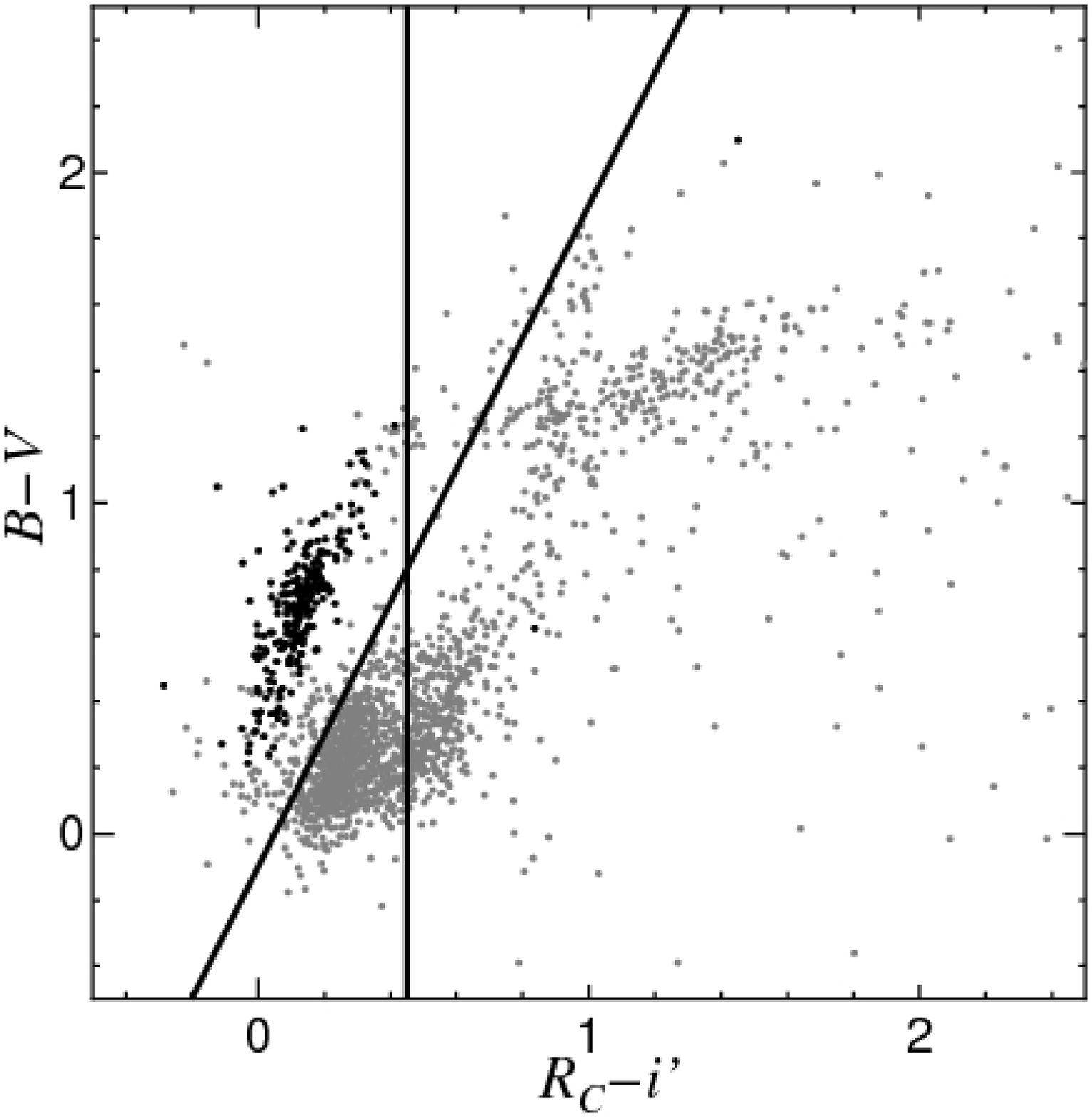}
\end{center}
\caption{
Diagrams of $B-V$ vs. $R_{\rm C}-i^\prime$. 
(Top) Colors of model galaxies (CWW) from $z=0$ to $z=3$ are shown with 
dotted lines: red, orange, green, and blue lines show the loci of E, Sbc, Scd, and Irr 
galaxies, respectively. 
Colors of $z=0.24$, $0.64$, $0.68$, and $1.18$
(for H$\alpha$, [O {\sc iii}], H$\beta$, and [O {\sc ii}] emitters, respectively) 
are shown with red, orange, green, and blue lines, respectively. 
Galaxies in the GOODS-N (Cowie et al. 2004) with redshifts corresponding to H$\alpha$ 
emitters, [O{\sc iii}] emitters, H$\beta$ emitters, and [O{\sc ii}] emitters are shown 
as red, orange, green, and blue open squares, respectively. 
Galaxies in the SDF (L07) with redshifts corresponding to H$\alpha$ 
emitters, [O{\sc iii}] emitters, and [O{\sc ii}] emitters are shown 
as red, orange, and blue filled squares, respectively. 
Solid lines show $(B-V) = 2(R_{\rm C}-i^\prime)-0.1$ and $(R_{\rm C}-i)=0.45$. 
The selection criteria for H$\alpha$ emitters by L07 are 
$(B-V) > 2(R_{\rm C}-i^\prime)-0.1$ and $(R_{\rm C}-i) < 0.45$. 
(Bottom) Plot of $B-V$ vs. $V-R_{c}$ for 2072 emission line candidates (gray filled circles) 
and 258 H$\alpha$ emitters (black filled circles).
\label{color3}}
\end{figure}

\begin{figure}
\begin{center}
\FigureFile(80mm,150mm){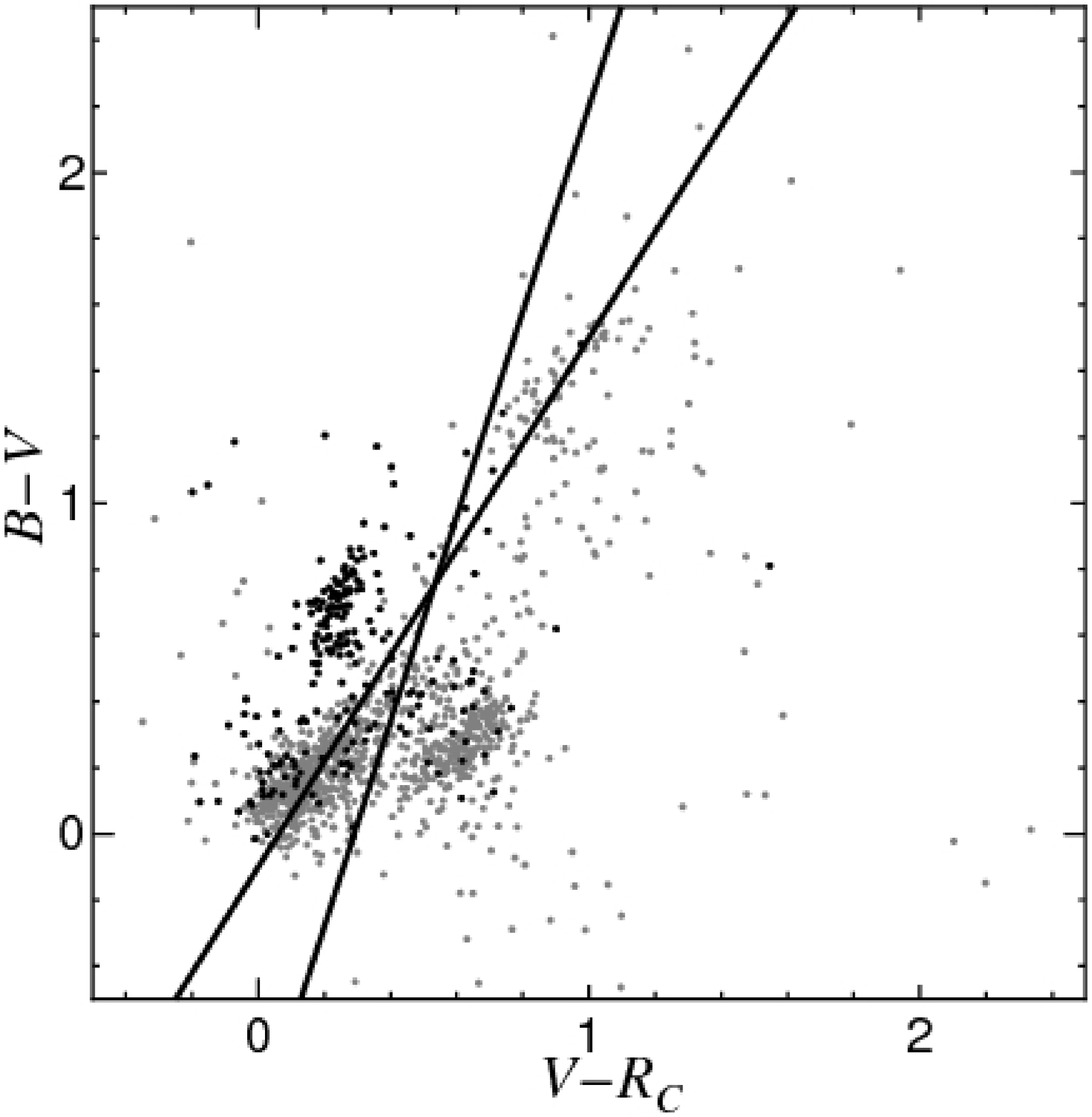}

\FigureFile(80mm,150mm){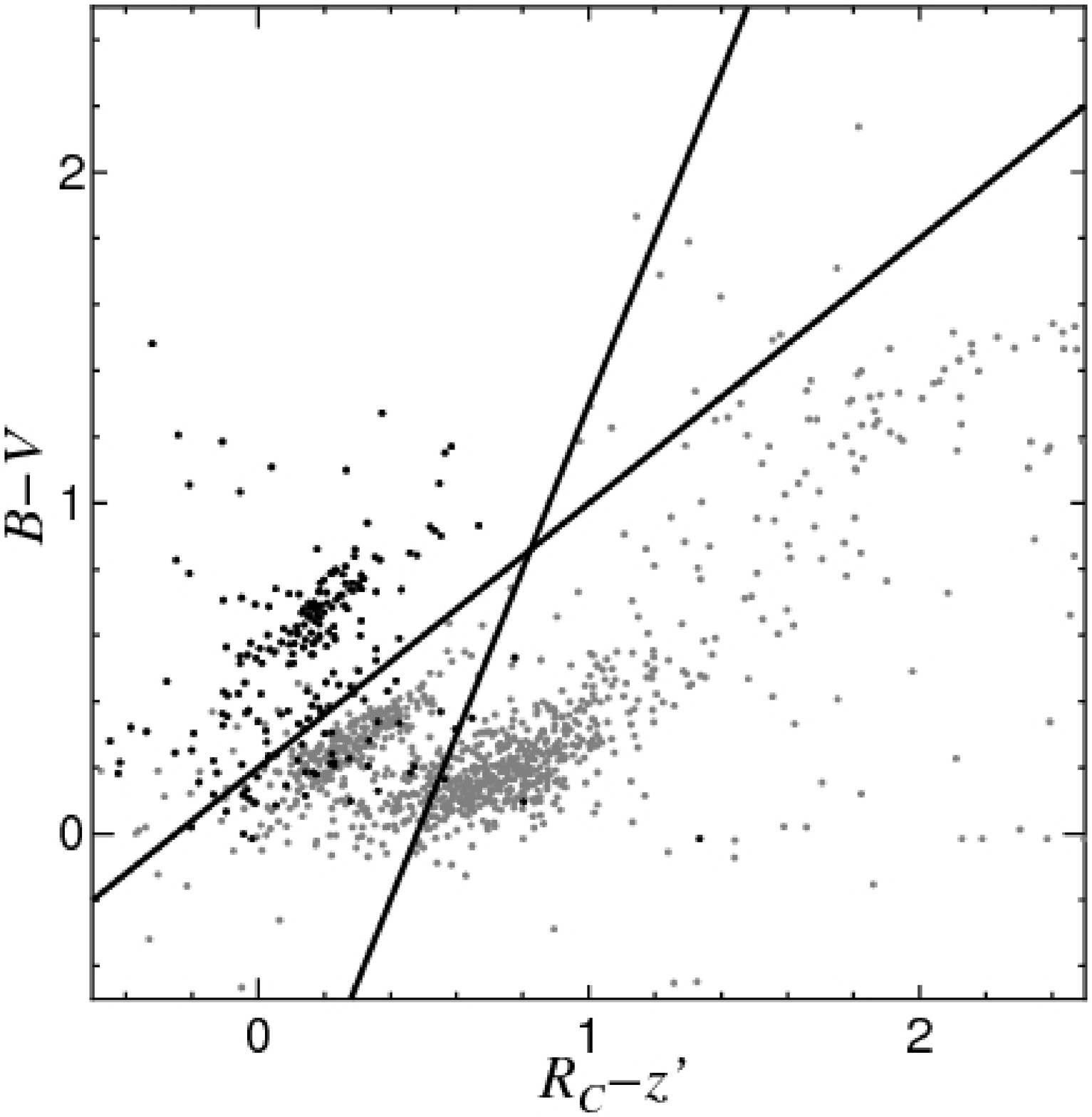}
\end{center}
\caption{
H$\alpha$ emitters using the L07
criterion selection rather than our new selection presented in Figures 2,3 \& 4. 
(Top) Plot of $B-V$ vs. $V-r^\prime$ for 
1345 emitter candidates (gray filled circles) and 
222 H$\alpha$ emitter candidates (black filled circles). 
(Bottom) Plot of $B-V$ vs. $R_{c}-z^\prime$ for 
1345 emitter candidates (gray filled circles) and 
222 H$\alpha$ emitter candidates (black filled circles). 
\label{color4}}
\end{figure}

Next, we compare our criteria for H$\alpha$ emitters with those of L07. 
We show the selection criteria of L07 and 
our emission-line galaxies and H$\alpha$-emitter candidates on 
the $B-V$ vs. $R_C-i^\prime$ plane in Figure \ref{color3}. 
Most of our H$\alpha$ emitters are satisfying the selection criteria of L07. 
We also plot emission-line galaxy candidates and H$\alpha$ emitters 
selected by the criteria of L07 on 
the $B-V$ vs. $V-R_{\rm C}$ and $B-V$ vs. $R_{\rm C}-z^\prime$ planes (Figure \ref{color4}). 
About one third of candidates do not satisfy our selection criteria. 
We consider that H$\alpha$ emitters of L07 may include [O{\sc iii}], 
H$\beta$, and [O{\sc ii}] emitters. 

\section{H$\alpha$ EMITTERS AT $z \approx 0.24$ IN THE SDSS}

In order to estimate a reliable bright-end slope of H$\alpha$LF,
we use a sample of brighter H$\alpha$ emitters taken from the SDSS Data Release 4 (DR4; Adelman-McCarthy et al. 2006)
adopting the same conditions of the SDF data; 
(1) $z$ = 0.233 -- 0.251 and (2) $EW$(H$\alpha$ + [N {\sc ii}] 
$\lambda\lambda$ $6548,6584$) $>$ 12 \AA.
To obtain a large sample of bright H$\alpha$ emitters, we use all SDSS DR4 data.
We then obtain a sample of 1371 H$\alpha$ emitters. 
It is noted that the obtained bright H$\alpha$ emitter sample is not taken from 
the SDF sky area. However, the spectroscopic area of DR4 is 4783 square degrees and 
thus the obtained bright H$\alpha$ sample provides a well-averaged one. 

\begin{figure}
\begin{center}
\FigureFile(150mm,150mm){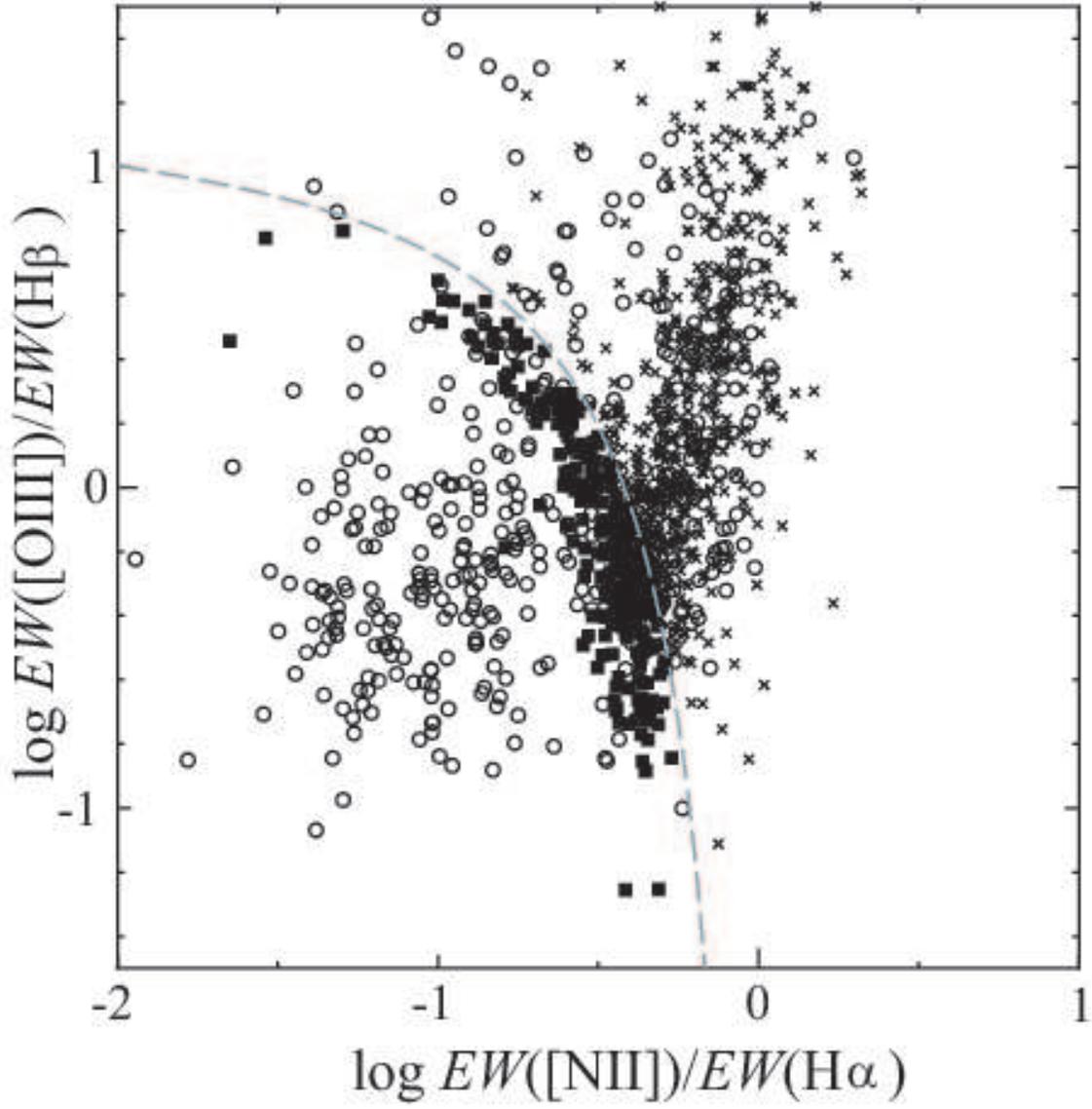}
\end{center}
\caption{Diagram between the $EW$ ratios of [O {\sc iii}]/H$\beta$ and [N {\sc ii}]/H$\alpha$ 
for all the galaxies in our SDSS sample. 
The gray dashed curve shows the dividing line between star forming galaxies and AGNs.
The samples are classified with star forming galaxies (filled squares), type 2 AGN (crosses), 
and type 1 AGN (open circles), respectively.
\label{BPTdiagram}}
\end{figure}

\begin{figure}
\begin{center}
\FigureFile(150mm,150mm){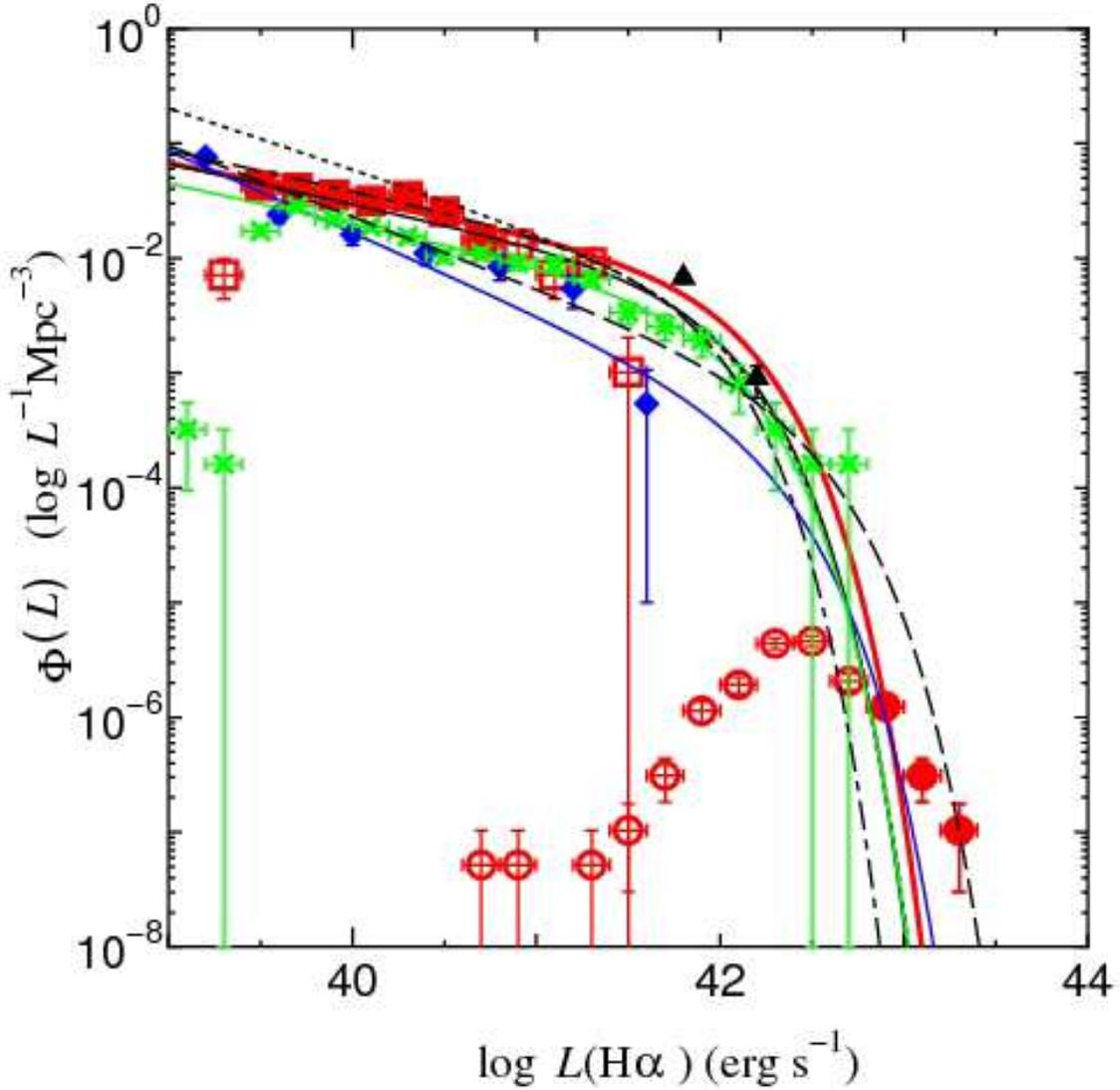}
\end{center}
\caption{
Luminosity function at $z\approx0.24$. 
Red marks and red solid line show our H$\alpha$ luminosity function. 
Squares and circles show the points derived from the SDF sample and the SDSS sample, 
respectively. 
Open symbols show the points we did not use when we fitted H$\alpha$ luminosity function 
with the Schechter function. 
The solid line shows the best fitted Schechter function. 
Blue diamonds and blue solid line show the H$\alpha$ luminosity function derived by L07. 
Green crosses and green solid line show the H$\alpha$ luminosity function derived by S08. 
Triangles show the H$\alpha$ luminosity functions derived by Pascual et al. (2001). 
The H$\alpha$ luminosity functions derived by Tresse \& Maddox (1998), F03, Sullivan et al. (2000), 
and Hippelein et al. (2003) are shown by black solid, black dotted, black dashed, and black dot-dashed line, 
respectively. 
\label{LF}}
\end{figure}

The SDSS H$\alpha$ emitter sample may contain not only star-forming galaxies but also 
type 1 and the type 2 AGNs.
Therefore, we have to exclude type 1 and type 2 AGNs from our initial SDSS sample.
We can reject type 1 AGNs from our sample automatically by using the SDSS spectral 
classification algorithm. In order to reject type 2 AGNs, we use the so-called emission-line
diagnostics (e.g., Baldwin, Phillips, \& Terlevich 1981; Veilleux \& Osterbrock 1987).
Figure \ref{BPTdiagram} shows a diagram between [O {\sc iii}]/H$\beta$ and 
[N {\sc ii}]/H$\alpha$ for our emission-line galaxies.
Usually their flux ratios are used in this kind of analysis. However, we use
$EW$ ratios in our analysis. Since the wavelengths of the two lines in the concerned
ratio are close, the $EW$ ratio can be used instead of the flux ratio. 
Following Kauffmann et al. (2003), a galaxy is defined as a type2-AGN if 
\begin{equation}
\log \frac{EW([\textrm{O {\sc iii}}])}{EW({\rm H \beta})} > \frac{0.61}{\displaystyle \log \frac{EW([\textrm{N {\sc ii}}]}{EW({\rm H \alpha})} - 0.05} + 1.3 \; .
\end{equation}
This curve is shown by the dashed line in Fig \ref{BPTdiagram}.

In this way, we can obtain our final SDSS sample of 317 H$\alpha$ emitters at $z \approx 0.24$
which are shown by filled boxes. These galaxies are considered to be 
star-forming galaxies.
In the SDSS sample, we estimate the H$\alpha$ luminosity in the same manner with SDF.
Since the limiting magnitude of the SDSS sample is $i^\prime = 19.1$,
the H$\alpha$ luminosity range is limited; $L({\rm H\alpha}) > 10^{42.8}$ ergs s$^{-1}$
for this SDSS H$\alpha$ emitter sample.

\section{RESULTS AND DISCUSSION}

\subsection{H$\alpha$ Luminosity Function at $z = 0.24$}

In order to investigate the star formation activity in galaxies at $z \approx 0.24$,
we construct the H$\alpha$LF by using the relation, 
$\Phi(\log L(\textrm{H}\alpha)) \Delta\log L(\textrm{H}\alpha) = \sum_i 1/V^i$,
where $V^i$ is the comoving volume, and the sum is over galaxies with H$\alpha$
luminosity within an interval of 
$\log L(\textrm{H}\alpha) \pm0.1 \Delta\log L(\textrm{H}\alpha)$.
It is unlikely that all the H$\alpha$ luminosity is produced by star formation
because active galactic nuclei (AGNs) can also contribute to the H$\alpha$ luminosity.
In the SDSS sample, we have distinguished the star-forming galaxies from AGNs
by using the excitation diagram (Fig \ref{BPTdiagram}).
However, in our SDF sample, we cannot apply the same method because 
no spectroscopic information is available.
We therefore adopt a statistical method following the manner of Pascual et al. (2001). 
The amount of AGNs depends on the selection criteria.
For example, it is 8-17\% in the CFRS low-$z$ sample (Tresse et al.
1996), 8\% in the local UCM (Gallego et al. 1995), and 17 -- 28\% in
the 15R survey (Carter et al. 2001) for the number of galaxies.
In the following analysis, we adopt that AGNs contribute to 8\% of the number density 
in the SDF H$\alpha$LF.

Combining the two H$\alpha$ emitter samples obtained from SDF 
and SDSS, we make a fitting with the Schechter function 
(Schechter 1976) using STY method (Sandage, Tammann, \& Yahil 1979) 
with error estimations adopting the method by Marchesini et al. (2007), 
and then obtain the H$\alpha$LF parameters; 
$\alpha = -1.31^{+0.17}_{-0.17}$, $\log \phi^{*} = -2.46^{+0.34}_{-0.40} {\rm Mpc^{-3}}$, 
and $\log L^{*} = 41.99^{+0.08}_{-0.07} {\rm ergs \; s^{-1}}$.
Our result is shown in Figure \ref{LF} 
together with H$\alpha$LFs of previous studies at $z < 0.3$ 
(Tresse \& Maddox (1998); F03; Hippelein et al. 2003; L07 and S08). 
We note that F03, L07, S08, and this work are based on 
the $\mathit{NB816}$ imaging obtained with the Subaru Telescope. 

First, we compare our H$\alpha$LF with that derived by L07. 
Although their H$\alpha$ luminosity was derived from the official catalog of the SDF as our work, 
their best-fit Schechter function parameters 
($\alpha = -1.71$, $\log \phi_*=-3.7$, $\log L_*=42.2$) 
are quite different from those of our H$\alpha$LF. 
Comparing our data points with those shown in Fig.10b of L07, 
we find that the slope of their LF at the faint end 
($\log L({\rm H}\alpha) \sim 39.5$-- 41.0) is similar to ours, 
although their number density is lower than ours by a factor of two.
This difference of the number density is explained by the different criteria of the sample selection: 
they selected emission line galaxies with larger $EW$s than ours. 
The origin of the significant difference of Schechter parameters 
between ours and L07's is considered that their fit is overweighted towards 
the faintest data points with the smallest error bars. 

Next, we compare our H$\alpha$LF with the other H$\alpha$LFs. 
Although our H$\alpha$LF is similar to those of Tresse \& Maddox (1998) and 
Hippelein et al. (2003), the H$\alpha$LF of F03 shows a steeper 
faint-end slope and a higher number density for the same luminosity than ours. 
L07 and S08 pointed out that the H$\alpha$ emitter sample of F03 may include [O{\sc iii}] 
or H$\beta$ emitters because of the inappropriate selection criteria of H$\alpha$ emitters. 
We consider this contaminant sources make the faint-end slope of H$\alpha$ luminosity 
function of F03 steeper than ours. 

\subsection{H$\alpha$ luminosity density and star formation rate density}

The H$\alpha$ luminosity density is obtained by integrating the H$\alpha$LF:
\begin{equation}
L(\textrm{H}\alpha) = \int_0^\infty \Phi(L) L\textrm{d}L = \Gamma (\alpha + 2) \phi^{*} L^{*}. 
\end{equation}
We then find a total H$\alpha$ luminosity per unit comoving volume
$4.45^{+2.96}_{-1.75} \times 10^{39} {\rm ergs \; s^{-1}}$ at $z\approx0.24$ from our
best fit H$\alpha$LF. 

The star formation rate (SFR) can be estimated from the H$\alpha$ luminosity using the relation, 
$SFR = 7.9 \times 10^{-42} L(\textrm{H}\alpha) ~ M_{\odot}$ yr$^{-1}$,
where $L(\textrm{H}\alpha)$ is in units of ergs s$^{-1}$ (Kennicutt 1998).
Thus, the H$\alpha$ luminosity density can be translated into the SFR density (SFRD) of
$\rho_{\rm{SFR}} \simeq 0.035^{+0.024}_{-0.014} ~ M_{\odot} \; {\rm yr^{-1} \; Mpc^{-3}}$ .
This value is higher by a factor of 3  than the local SFRD.

\begin{figure}
\begin{center}
\FigureFile(150mm,150mm){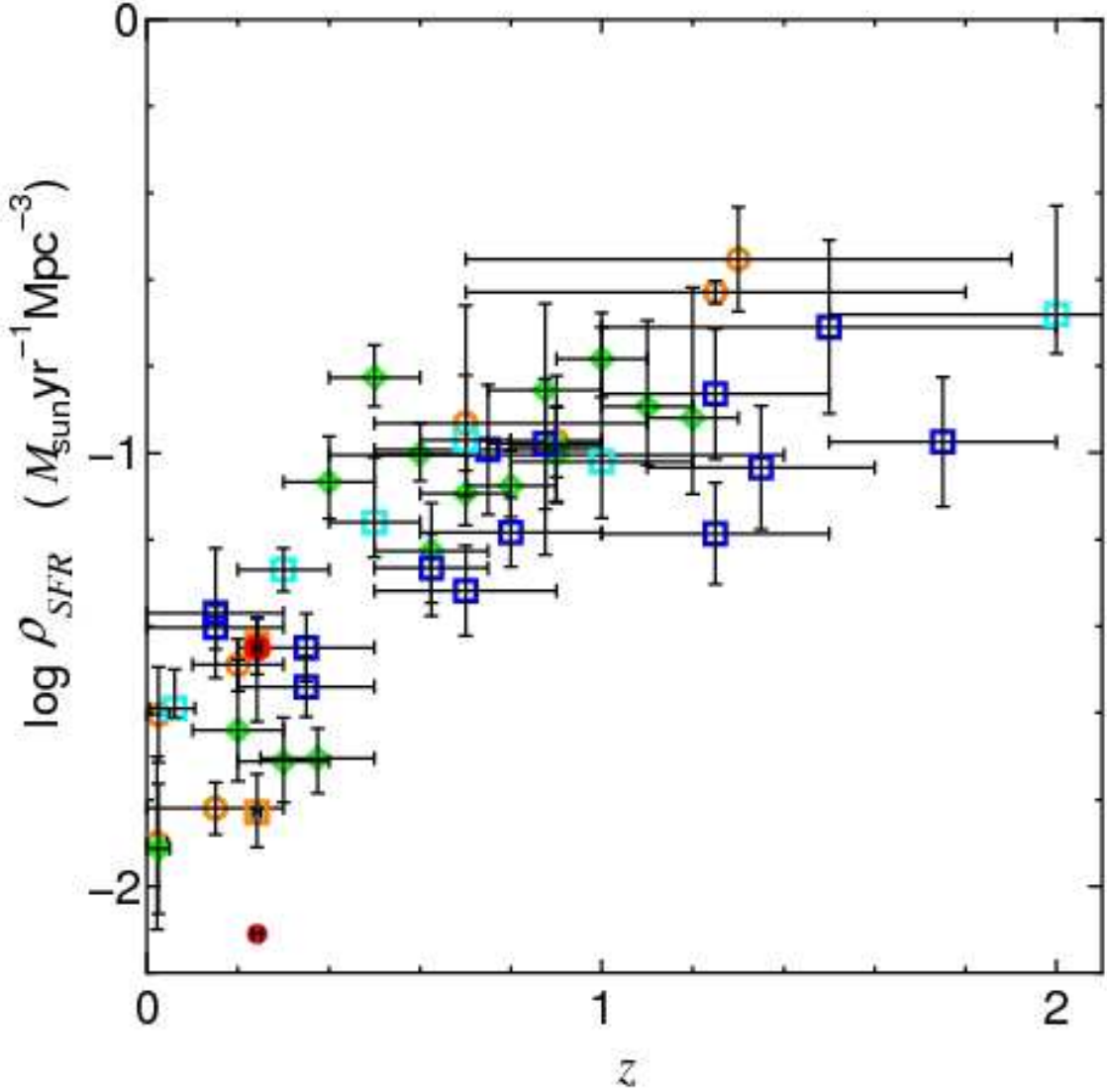}
\end{center}
\caption{
Star formation rate density (SFRD) at $z\approx0.24$ derived from 
our study (the large red filled circle) shown together with the previous 
investigations compiled by Hopkins (2004). 
SFRD estimated from H$\alpha$, [O {\sc ii}], and UV continuum are shown 
with orange open circles (P\'erez-Gonz\'alez et al. 2003; Tresse et al. 2002; 
Moorwood et al. 2000; Hopkins et al. 2000; Sullivan et al. 2000; Glazebrook et al. 1999; 
Yan et al. 1999; Tresse \& Maddox 1998; Gallego et al. 1995), 
green open diamonds (Teplitz et al. 2003; Gallego et al. 2002; Hogg et al. 1998; 
Hammer et al. 1997), 
and blue squares (Sullivan et al. 2000; Connolly et al. 1997; Lilly et al. 1996). 
The light blue open squares show SFRDs based on the UV luminosity density by 
Schiminovich et al. (2005), assuming $A_{\rm FUV}=1.8$. 
Small orange filled circle shows SFRD at $z \approx 0.24$ derived 
by F03. 
Orange open square shows SFRD derived for the COSMOS field (S08). 
The SFRD derived by L07 is shown as small red filled circle. 
\label{SFRD}}
\end{figure}

We compare our result with the previous investigations compiled by Hopkins (2004) 
in Figure \ref{SFRD}. 
We also show the evolution of the SFRD derived from the observation of GALEX 
(Schiminovich et al. 2005) with mean attenuation of $A_{\rm UV}^{\rm meas} = 1.8$, 
evaluated from the relation between $A_{UV}$ and FUV slope for starburst galaxies (Meurer et al. 1999).  
Recent observations showed that the typical value of $A_{UV}$ is 1 mag for nearby galaxies 
detected in the UV by GALEX (Seibert et al. 2005; Buat et al. 2005; Tresse et al. 2007). 
If we adopt the attenuation of $A_{\rm UV} = 1$, the SFRD derived from the FUV becomes 
smaller by factor 2. 

Our SFRD estimated above seems roughly consistent with the previous 
evaluations, e.g., Tresse \& Maddox (1998) and F03. 
We note that the SFRD of F03 
was overestimated because of the contamination of [O{\sc iii}] emitters. 
The SFRD of S08 is slightly smaller than ours. 
Since our selection criteria are similar to those of 08, 
the difference between S08 and our work seems to be real; 
e.g., the cosmic variance. 
The SFRD of L07 is smaller than other evaluations. 
The small SFRD of L07 is originated from the different 
selection criterion of H$\alpha$ emitters as we mentioned in section 4.1. 

\subsection{Spatial Distribution and Angular Two-Point Correlation Function in the SDF}

\begin{figure}
\begin{center}
\FigureFile(150mm,150mm){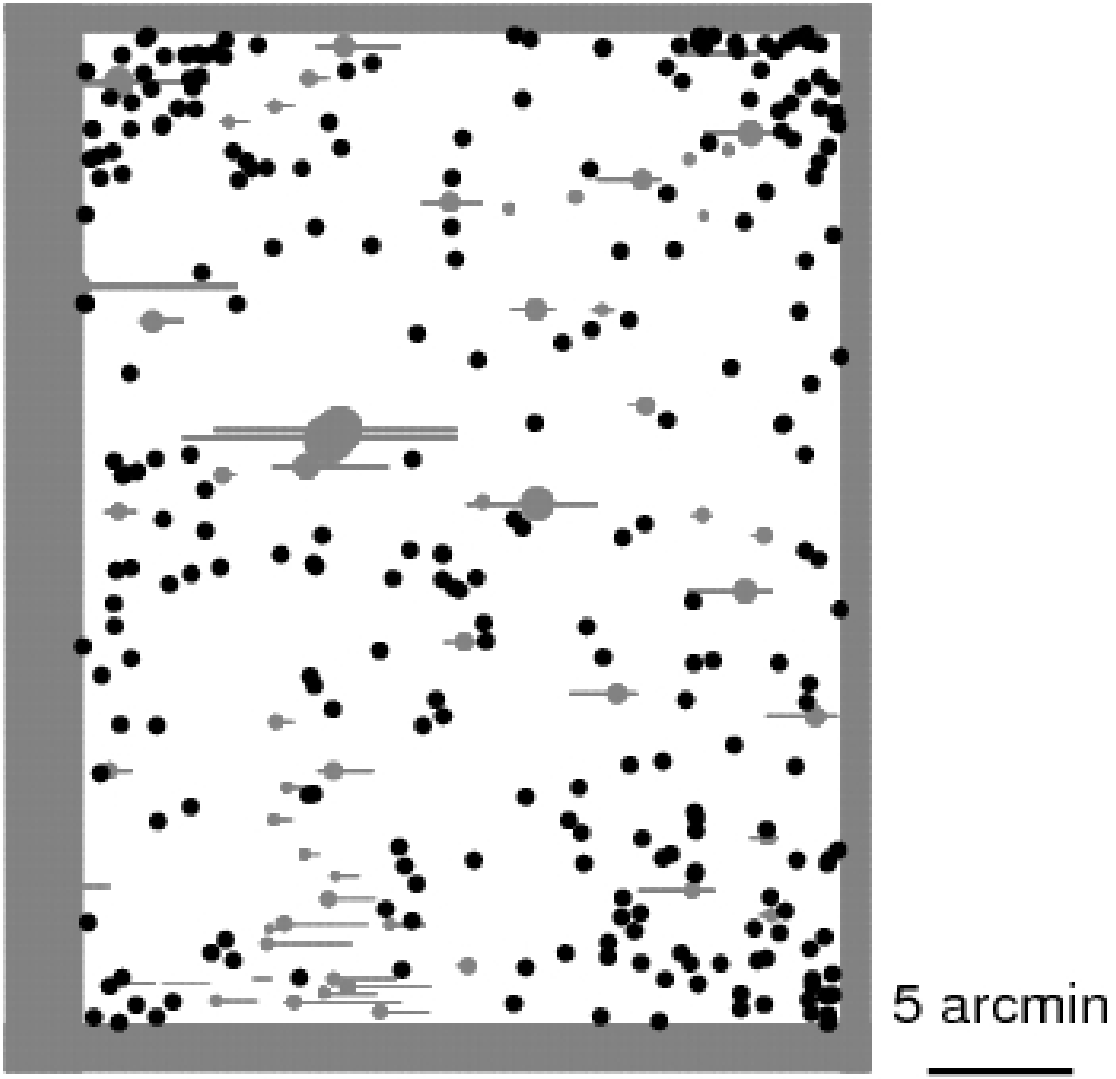}
\end{center}
\caption{
Spatial distribution of our H$\alpha$ emitter candidates. 
Shadowed regions show the areas masked out for the detection.
\label{XY}}
\end{figure}

\begin{figure}
\begin{center}
\FigureFile(150mm,150mm){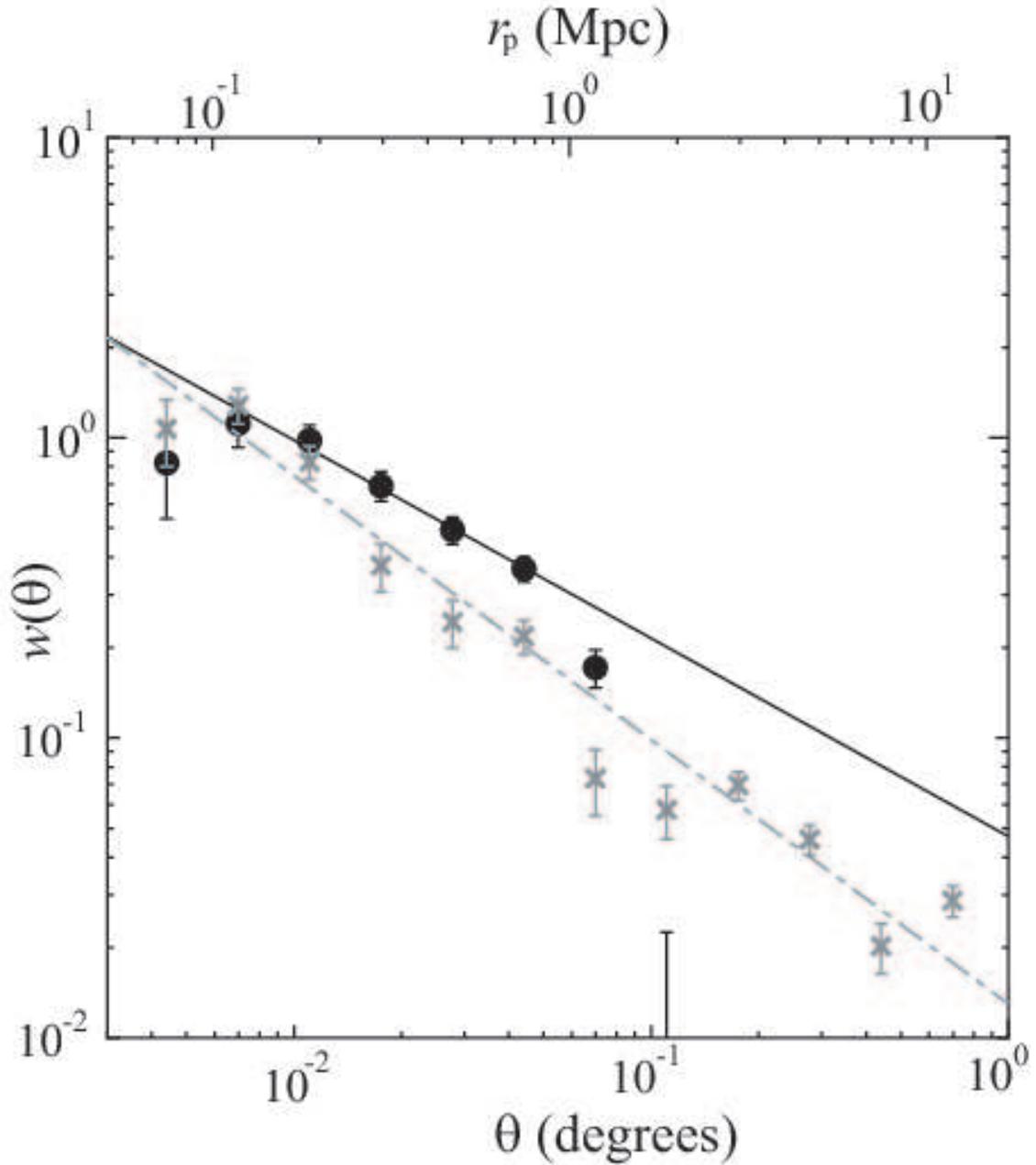}
\end{center}
\caption{
Angular two-point correlation function of all H$\alpha$ emitter candidates (filled circles). 
Solid line shows the relation of $w(\theta) = 0.047 \theta^{-0.66}$. 
For comparison, we also plot the angular two-point correlation function of 
H$\alpha$ emitters in COSMOS (gray crosses; S08). 
Gray dot-dashed line shows the relation of $w(\theta) = 0.013 \theta^{-0.88}$ (S08). 
\label{ACF}}
\end{figure}

We show the spatial distribution of our 258 H$\alpha$ emitter candidates in Figure \ref{XY}.
There are some concentrations of H$\alpha$ emitters in the corners. 
To discuss the clustering properties more quantitatively, 
we derive the angular two-point correlation function (ACF), $w(\theta)$,
using the estimator defined by Landy \& Szalay (1993),
\begin{equation}
 w(\theta) = \frac{DD(\theta)-2DR(\theta)+RR(\theta)}{RR(\theta)},
 \label{two-point}
\end{equation}
where $DD(\theta)$, $DR(\theta)$, and $RR(\theta)$ are normalized numbers of
galaxy-galaxy, galaxy-random, and random-random pairs, respectively.
The random sample consists of 100,000 sources with the same geometrical
constraints as the galaxy sample.
The formal error in $w(\theta)$ is described by 
\begin{equation}
\sigma_w=\sqrt{[1+w(\theta)]/DD}
\end{equation}
(Hewett 1982). 
We show the ACF for all our 258 H$\alpha$ emitter candidates in Figure \ref{ACF} 
together with that for 980 H$\alpha$ emitters in the COSMOS field (S08).
The FOV areas from which we derive $w(\theta)$ are $37^\prime \times 27^\prime$ for 
the SDF and $1.4^\circ \times 1.4^\circ$ for the COSMOS field. 
The amplitude of our ACF is positive within the 0.06 degree ($\sim 4$ arcmin) 
and it is larger than that of the COSMOS field. 
We consider that this clustering property reflects 
the concentration of H$\alpha$ emitters in the corners. 
Our ACF is fit well by power law, $w(\theta) = 0.047^{+0.017}_{-0.013} \theta^{-0.66 \pm 0.08}$. 

It is useful to evaluate the correlation length $r_0$ of the two-point 
correlation function $\xi(r) = (r/r_0)^{-\gamma}$. 
A correlation length is derived from the ACF through Limber's equation 
(e.g., Peebles 1980). 
Assuming that the redshift distribution of H$\alpha$ emitters is 
a top hat shape of $z=0.242 \pm 0.009$, we obtain the correlation 
length of $r_0 = 2.6^{+1.0}_{-0.8}$ Mpc. 
Therefore, the two-point correlation function for all H$\alpha$ emitters 
is written as $\xi(r) = (r/{\rm 2.6^{+1.0}_{-0.8}Mpc})^{-1.66 \pm 0.08}$. 
This correlation length is larger and the slope of the correlation function is shallower 
then those evaluated for H$\alpha$ emitter at $z \approx 0.24$ in COSMOS; 
$\xi(r) = (r/{\rm 1.9Mpc})^{-1.88}$ (S08). 
We note that the correlation lengths referred in this section are evaluated for the cosmological 
parameters of $\Omega_{\rm matter}=0.3$, $\Omega_\Lambda=0.7$, and $H_0=70 {\rm km \; s^{-1} \; Mpc^{-1}}$. 
This difference reflects the difference of the amplitude of the ACF within $\sim 4$ arcmin. 
The correlation length is smaller than that of nearby $L_*$ galaxies 
($\sim 7$ Mpc; Loveday et al. 1995; Norberg et al. 2001; Zehavi et al. 2005) 
and $z \sim 1$ galaxies ($\sim$ 4--5 Mpc; Coil et al. 2004)
It is known that the correlation length is smaller for fainter galaxies 
and is smaller for later-type galaxies in the nearby universe 
(Loveday et al. 1999; Norberg et al. 2001; Zehavi et al. 2005). 
Although the absolute $R$-band magnitudes of our sample ($M_R \sim -18$) is faint, 
the correlation length is still smaller than the nearby galaxies with the same absolute magnitude 
(3.8 Mpc; Zehavi et al. 2005). 
The clustering strength of emission-line galaxies (active star-forming galaxies) may be 
weaker than that of non-active star-forming galaxies. 

\vspace{1pc}
We would like to thank the Subaru Telescope staff for their invaluable help. 
We would like to thank Nobunari Kashikawa who is the PI of SDF,
and the other SDF members. 
We also thank Chun Ly and Matt Malkan who gave us useful information of their 
independent study of H$\alpha$ emitters in SDF prior to the publication. 
This work has been financially supported in part by grants of JSPS (Nos. 15340059
and 17253001). 


\clearpage


\begin{table}
\caption{Emission lines potentially detected inside the
narrowband.}\label{Ha:tab:cover}
\begin{center}
\begin{tabular}{lcccc}
\hline
\hline
{Line} &
{Redshift range} &
{$\bar{z}$\footnotemark[$\dagger$]} &
{$d_L$} &
{$V\times10^{4}$\footnotemark[$\ddagger$]} \\
{} &
{$z_1\leq z \leq z_2$} &
{} &
{(Mpc)} &
{(Mpc$^3$)}\\
\hline
H$\alpha$+[N{\sc ii}]6548,6584  & 0.233~~~0.251 & 0.242 & 1220  & 0.49\\
{[O {\sc iii}]} $\lambda \lambda$ 4959,5007 & 0.616~~~0.640 & 0.628 & 3740  & 2.96\\
H$\beta$  & 0.664~~~0.689 & 0.677 & 4100  & 3.25\\
{[O {\sc ii}]} $\lambda$ 3727   & 1.17 ~~~ 1.20 & 1.19  & 8190  & 9.05\\
\hline
\multicolumn{4}{l}{\footnotemark[$\dagger$] {Mean redshift.}}\\
\multicolumn{4}{l}{\footnotemark[$\ddagger$] {Comoving volume.}}\\
\end{tabular}
\end{center}
\end{table}


\end{document}